\documentstyle[12pt,prepriop]{article}
\begin{document}
\begin{center}
\bf   THE DYNAMICS OF VORTEX STRUCTURES AND STATES OF CURRENT IN 
PLASMA-LIKE FLUIDS AND THE ELECTRICAL EXPLOSION OF CONDUCTORS:\\
3. Comparison with Experiment.
\end{center}
\begin{center}
N B Volkov and A M Iskoldsky\\
Russian Academy of Science, Ural Division\\
Institute of Electrophysics\\
34 Komsomolskaya St.\\
Ekaterinburg 620219,\\ 
Russia
\end{center}
\vspace{3.5cm}
\begin{center}
\bf The Dynamics of Vortex Structures and States of Current: 3
\end{center}
\newpage
\begin{center}
{\bf THE DYNAMICS OF VORTEX STRUCTURES AND STATES OF CURRENT IN 
PLASMA-LIKE FLUIDS AND THE ELECTRICAL EXPLOSION OF CONDUCTORS:\\    
 3. Comparison with Experiment.}\\ 
                     N B Volkov and A M Iskoldsky
\end{center}
\begin{center}
\bf Abstract
\end{center}
The present paper is a concluding part of the series [N~B~Volkov
and A~M~Iskoldsky: 1; 2; [1],[2]]. Here on the basis of the 
results of the 
above papers the experiments on the electric explosion  in  conductors 
have been discussed and the mechanism of "hot" plasma spots  formation 
which was repeatedly observed in the experiments on  a  plasma  focus, 
z-pinches, vacuum sparks and electric explosion of microwires has been 
proposed. To realize this mechanism,  the  formation  of  sausage-type 
instabilities is not necessary. Moreover, the possibility of hot spots 
formation in a liquid metal (in the model under discussion, the
plasma-like medium is assumed incompressible and the local
kinetic transport coefficients - constant) has been shown. The
latter is of great 
interest for understanding the mechanism of initiation  and  operation 
of vacuum arc cathode spots  and  also  for  the  explanation  to  the 
phenomenon of explosive electron emission which is widely employed  in 
applied physics. 
\newpage
\section{Introduction}
In the present paper which is a third part of the series 
[1], [2] we confine ourselves to a comparison with old 
experiments on the 
electric explosion of conductors (EEC), which represent 
a large  volume  of experimental data [3], [4], 
[5]. They contain  a  considerable  number  of 
facts, known as anomalies of electric explosion, which have not  found 
a satisfactory interpretation so far. Close to this problem are papers 
on liquid-metal current interrupters, Z-pinches 
[6] and  plasma  focus experiments 
[7], and plasma-erosion current breakers [8]. 
\section{Experimental techniques}
\par
{\bf 2.1.} For a more objective discussion of the experimental data  on 
EEC, let us review experimental technique  (the  problem  has  a  more 
detailed consideration in [3], [4], [5]). The major 
experiments discussed 
below were carried out with  the  RLC  circuit  corresponding  to  the 
circuit S3 in [2] without the load resistor 
$R$ (Fig.1c), on  the  basis 
of high-voltage ($U_C = 50~kV$), low-inductive capacitors with the  total 
capacity $(2.5 - 8)~{\mu}F$. The initial charge voltage 
ranged from 10 to 50~$kV$ and the period of oscillations was 
between 5 and 40~${\mu}s$. The maximum 
current amplitude was limited by 100~$kA$. The accuracy of the 
initial charge voltage was about $10^{-3}$. In most experiments 
the initial parameters were adjusted so that the stage of the 
conductor melting began 
within the maximum of the first half-cycle of a current wave in 
the RLC circuit. A typical value of the current density 
($\bi{j}$) for this  type of source is about $2\cdot10^7~A{cm}^{-2}$. 
\par
{\bf 2.2.} A three-channel square-wave generator made of 
50-Ohm, rf cables [9] was used to obtain maximal current 
densities of $(1 - 5)\cdot10^8~A{cm}^{-2}$. The pulse rise 
time was 1~$ns$.  One of the generator 
channels was connected to the load and the other two 
served to control the electron-optical image converter. 
In the capacity of the load the authors [4] 
and [9] used a microwire of 10 microns 
in diameter and several millimeters in length. 
\par
{\bf 2.3.} The experiments mostly dealt  with two observables: 
the  total 
current, measured with the help of a special shunt, and the 
voltage, 
measured by a compensated Ohmic divider installed parallel to 
the 
conductor. The signals were recorded by rapid oscilloscopes 
with a bandwith of 3~$GHz$ or a rapid double-channel digitizer 
with the 
sampling time 7~$ns$ and the amplitude resolution $1\%$. 
\par
{\bf 2.4.}  Furthermore, images of the  exploding  conductor 
were  taken  by  an 
electron-optical and by a pulse shadow X-ray photography. The  picture 
quality in X-ray photographs obtained in [4] 
is characterized  by  the 
following figures: the unsharpness of the image edge is of  the  order 
of 20 microns, the number of allowed shading  is  more  than  10,  the 
minimal allowed thickness is of the order of 50 microns of Al and  the 
maximal one is of the order of 1~$mm$ of  Cu,  the  accuracy  of  timing 
photographs to voltage waveforms is within  $2\cdot10^{-8}~s^{-1}$. 
\par
{\bf 2.5.} Effective methods of obtaining and processing the 
experimental data allowed us to find out insignificant (of the 
order of a few percent) changes in the data caused by physical 
fluctuation processes. For 
instance, registered with a high level of significance was a 
change within $10\%$ in the graph slope of the resistivity 
against 
the absorbed energy which occurred in the stage preceding 
melting and also under similar 
conditions, a shift of coordinates of the  initial  melting  point  by 
about $10\%$, ets. [4]. 
\section{Discussion} 
In most experiments (see [3],
[5]) only two signals are often measured: the 
total current and voltage drop across the conductor. Therefore, 
below we basically discuss experiments of [4], where in addition 
to the measurements mentioned above the shadow X-ray photography and the 
photography by an electron-optical image converter were used. 
\par
{\bf 3.1.} Figure 1 shows typical waveforms of the current 
(Fig. 1a) and the voltage drop (Fig. 1b) without an  inductive 
component. Shown in Fig. 1c is the  time  variation  of  the  relative 
conductor radius taken by the shadow X-ray  photography.  The  initial 
point of an electric explosion on the  current  curve  is  practically 
indistinguishable though on the voltage  curve  it  is  marked  by  an 
increase in the voltage growth rate. However the transition region  is 
sufficiently wide. Therefore it is not simple to determine the initial 
explosion point by the  current  and  voltage  waveforms.  Independent 
measurements of the conductor radius  during  the  explosion  enables 
removal of the ambiguity (the change in  the  mechanism  of  conductor 
expansion takes place at the initial explosion point (Fig. 1c)).
\par
The ratio of a voltage signal to the current is interpreted as the 
conductor resistance, their product - as the absorbed  power  and  the 
integral of this product over time -  as  the  absorbed  energy.  This 
interpretation is based on a priori assessments  of  a  characteristic 
time of the magnetic field diffusion. This time is shorter 
than the characteristic time of the change in current and for 
these reasons 
the current density as well as other  parameters  are  assumed  to  be 
evenly distributed over the conductor volume. By using the  dependence 
on time and absorbed energy found  in  this  way,  one  can  determine 
parameters corresponding to the initial explosion point: the 
time ($t_c$) and the absorbed energy $Q(t_c)$. 
\par
The latter quantities have been widely discussed in the literature 
[3]. The authors of [3], 
[10] state that $Q(t_c)$ is independent of the 
current density, that is of the conductor heating rate, if the 
conductor is 
homogeneous over the radius $(t_c\gg{r_0}{c_s}^{-1}$, $r_0$, 
$c_s$ 
are  the  conductor 
radius and the sound velocity respectively).  Moreover,  they  suggest 
defining the initial explosion point as a point corresponding  to  the 
same resistance of a conductor made of the same material (for 
$W$, e.g., the resistance ($R_{eff}$) with $t = t_c$ equal to 
the resistance at the end 
of melting is chosen in [10]), or as a point 
corresponding to the time 
when the condition $v(t ){v_0}^{-1}$ = 1.35 - 1.45 ($v$ and 
$v_0$ are  the instantaneuos  
and initial volumes of the conductor respectively) is 
satisfied [3]. This estimate calculated for the 
radius has the form $r_c = r(t_c) = (1.162 - 1.204)r_0$. The 
lower estimate is in good agreement with 
experimental results of [4] (Fig. 1c). 
\par
In order to check the hypothesis of $Q(t_c)$ dependence on the 
current density and the linear dependence of $R_{eff}(Q)$ in 
the 
section b - c (Fig.1), we simulated this situation in the 
circuit S2 of our paper [2] where we calculated 
the process with two values of the 
dimensionless load impedance ($\Pi_3$): $\Pi_3 = 1$ and 
$\Pi_3 = 4$ (Fig. 2a, b, c). It can be seen that $R_{eff}(Q)$ 
in this section may be really approximated 
by the linear dependence, the resistance corresponding to a 
higher energy input rate into the conductor being in the region 
of a high 
absorbed energy. From the results of our calculations, we can  suggest 
the following about the parameters of the initial explosion point.  If 
similar to [3], [10], we assume 
$Q(t_c)$ independent of the energy input rate, then for the 
above value we should take a point with $Q = 1.71\cdot10^{-1}$ 
and $t\cong10.1$ in Fig. 2a (the units of measurement are 
relative), which is located far from the region of a sharp increase in 
$R_{eff}$. According to Fig.  1c, however, the initial 
explosion point 
localizes at the beginning of the region of a fast $R_{eff}$ 
increase. In analyzing Fig. 2 one should bear in mind that in 
our model the fluid is considered to be incompressible and 
kinetic coefficients to be constant.  In 
addition, the growth of $R_{eff}$ is determined not by the 
dependence of 
electric conductivity on temperature but by  the  dynamics  of  vortex 
states, which results in the fact that an unperturbed  state  with  an 
evenly distributed  current  density  over  the  conductor  radius  is 
essentially disturbed. As a result, it is not only the  
current  density  that  is 
unevenly distributed over the conductor but the temperature  as  well. 
Due to this fact one of the anomalies of  the  electric  explosion  is 
removed, namely the problem of the conductor overheating (see  below). 
Moreover,  the  interpretation  of  experiments  based  on  a   priori 
assessments of a short diffusion time can be assumed to be misleading. 
\par
{\bf 3.2.} An essential feature of the electric explosion in conductors 
is their critical (threshold) character. The  model  discussed  has  a 
natural criterion (the magnetic Rayleigh number) and the  question  of 
whether the explosion transition has to occur is solved positively, at 
least for the source of an infinite power, if this parameter surpasses 
a critical value defined by the geometry and the conductor properties. 
For the source with a finite energy content, as pointed in [2], the 
situation is more complex. In this case, the question of  whether  the 
explosion has to occur is determined not only by the control parameter 
$r_1$, which equals the ratio of the magnetic Rayleigh number to the 
critical one, but also by the initial level of fluctuations in  the 
system given by the amplitudes $X(t)$, $Y(t)$ and $Z(t)$ ($X$ belongs to 
mechanical degrees of freedom and $Y$ and $Z$ - to electromagnetic  ones). 
Our model is applicable from the time $t_0$ which corresponds to the  end 
of the conductor melting. The experiment in 
[4] shows that the initial 
and final melting points shift towards the region of a  high  absorbed 
energy where the values of $R_{eff}$ and the slope angle 
$R_{eff}(Q)$ are smaller 
than the table values. Moreover, the same experiment shows that 
the resistance of a wire made of such materials  as  steel,  tungsten  and 
molybdenum decreases with increasing $Q$. Therefore when  comparing  our 
model with the experiment, there arises a problem of not only to  find 
a critical value of the control parameter $r_1$ but also  to  define  the 
initial data. It can be easily shown that by choosing the value of $Z_0$, 
one can obtain a decrease in $R_{eff}$ and by choosing 
$X_0$ - ashift towards the region of a high absorbed energy. 
\par
{\bf 3.3.} Experiments in [11] have shown that if the 
current is rapidly 
interrupted before the conductor is brought up to its  melting  point, 
then the conductor is destroyed after about 200~${\mu}s$. In a similar  way, 
when the current is  interrupted  before  the  conductor  reaches  its 
melting point, the conductor starts bending in  some  delay  time.  In 
[12], for a Cu conductor 1~mm in diameter, the 
delay time was about 100~${\mu}s$. In [4], 
bending in the central section of the conductor forms 
when the absorbed energy corresponds to that for liquefaction 
of the metal.  This 
indicates that on the one hand, the wave length of the instability due 
to bending is fixed apparently when the conductor  material  is  elastic 
and on the other hand, that the melt can  resist  shearing  stress  at 
least during a few microseconds after it has been produced. 
In terms  of  our 
model (see [1]), the formation of vortex 
hydrodynamical structures 
introduces elements of a solid into the fluid:  an  effective  voltage 
tensor is non-diagonal and  the  fluid  gains  in  ability  to  resist 
deformation. It should be noted that vortex excitations are similar to 
disclinations in liquid and solid crystals. Therefore the  development 
of  calibration  continuous  models  of  defects   (dislocations   and 
disclinations), which interact with the electromagnetic field  of  the 
model type proposed in [13], represents an 
up-to-date problem in the study of 
pulse heating and melting of conductors exposed to heating by current. 
These models will enable not only the explanation of destruction after 
the interruption of current but also the solution of a problem of  the 
starting level of disbalance for the liquid metal stage. 
\par
{\bf 3.4.} One of the most indicative features of the electric 
explosion is the emergence of a characteristic large-scale 
bandlike structure 
or strata dividing the conductor into lengths of a typical size  close 
to the initial radius of the conductor (Fig. 3 (see  also  Fig.  2  in 
[1])). Shown in Fig. 3 are the data of 
explosion for a Cu conductor 
0.58~$mm$ in diameter in an electric circuit with the period 
40~${\mu}s$, the 
storage capacity 4.2~${\mu}F$ and the initial voltage 30~$kV$. 
The voltage 
rise time from the starting point of  the  explosion  to  the  maximum 
value (points "c" and "e", Fig  1a)  was  1.2~${\mu}s$.  The  fast  current 
interruption phase lasted 1 ms and the time of the second phase  (from 
the inflection point on the leading edge (Fig. 1a, point "d")  to  the 
point "e") was about 200~$ns$. The first picture of  Fig.  3  shows  the 
beginning of the first phase, the second picture was taken at its  end 
and the third one - near the maximum of  the  voltage  pulse.  In  the 
first picture one  can  see  an  axially  asymmetrical,  small-scale 
structure with a typical size of about 0.3~$mm$. In the second  picture, 
at least in three locations, plasma strata have been  already  formed. 
The  third  picture  is  a  typical  pattern  of  a  fully  stratified 
conductor. The mean distance between the strata is 0.78~$mm$, and it  is 
twice as large as a small-scale structure  at  the  beginning  of  the 
first phase. In addition, it appears that  the  typical  size  in  the 
first approximation is independent of the experimental condition,  and 
the time to form this structure may be shorter than the typical  sound 
time (it follows from the comparison of the last two pictures in  Fig. 
3, that the stratification process of the experiment discussed  lasted 
200~$ns$). 
\par
In the framework of our model the structure size  is  set  by 
the wave number of the most rapidly increasing harmonic, and in 
accordance with the analysis in [1] for a Cu 
conductor 0.58~$mm$ in diameter, 
it has to be $l = 2{\lambda} = 0.672~mm$. In the experiment 
from [4] discussed 
above, we have a value very close to $l = 0.78~mm$.  If we 
employ the 
radius value corresponding to the initial explosion point from 
[3] $r(t_c)\cong1.162{r_0}$, close to the 
experimental data from [4], we find a 
better agreement with experiment: $l = 0.781~mm$. 
Moreover, in later 
phases of the instability development one can observe the  coalescence 
of adjoining strata into pairs (see Fig. 2 from [1]). 
\par
Such fine effects are predicted  by  the  model:  the 
harmonic with the wave number $0.5k$, according to Fig. 1 from 
[1] has to 
be excited by the following one. To have an understanding 
of  the  pairing 
effect discussed, there is a well-known result of  the  vortex  system 
dynamics: two vortex rings are either attracted or repelled (depending 
on the relative orientation of the particle rotation near the axis  of 
vortex rings) [14]. The contact point between two 
repelling vortices can move from the external surface deep into 
the interior at  a  speed  well  above 
that of a sound, for this movement is a phase one (in 
[15] we developed a three-mode model of this 
stage of the process, which is similar to the model discussed 
here and where the $X$, $Y$ and $Z$ amplitudes 
have the following asymptote: $X{\sim}Y{\sim}Z{\sim}(t_* - t)^{-1/2}$). 
\par
{\bf 3.5.} At the beginning of the conductor  destruction,  the  energy 
deposited in it is sufficient to heat the conductor up to the 
temperature well above (by 2 - 3 times) the boiling point on 
the assumption 
that the temperature distribution is practically uniform over  volume. 
The experimental lifetime of  this  state  is  by  several  orders  of 
magnitude higher than the most  radical  estimations  derived  from  the 
theory (e.g., by Frenkel-Zeldovitch-Folmer [16]). 
The difficulty 
mentioned is absent from the model under discussion: in our 
case, because the 
temperature distribution over  the  conductor  volume  is  essentially 
nonuniform. The maxima of temperature can be well above the mean value 
and the surface temperature can be well  below  the  mean  value  over 
volume (see Fig. 4a). In our case, $T_{max} = 
23.43\langle{T}\rangle$ ($\langle\cdots\rangle$ means the 
volume averaging). 
\par
The maxima of X-ray intensity corresponds to the local 
temperature maxima (Fig. 4b; marked in the current curve 
(Fig. 4c) is the time for the plots in Fig. 4a,b). 
These regions of high temperature are  detected 
by a pinhole chamber and interpreted as the so-called hot plasma spots 
(see [17] and also [18], [19], [20]). The generally accepted 
theory assumes 
that hot spots form as a result of sausage-type instabilities 
(micro-pinches) caused by radiation cooling and the escape of 
particles 
of matter from the region of the sausage-type instability 
[21]. In our 
case, the self-focusing of current occurs not due to  "raking  up"  of 
matter but owing to the decrease in the cross-section of the  channel, 
along which a laminar component of current flows (see Fig. 9 
from [2] 
which shows cross-sections of separatrix  surfaces  in  the  field  of 
current velocities of conduction electrons). 
\par
We think that in the known experiments on micro-pinches 
[18], [19], [20], 
the discussed self-focusing of current is also realized. Our ideas are 
supported by experiments of [19], carried out 
with the help of the 
X-ray electron-optical image converter with the time resolution  about 
5~$ns$, which found (by direct observation) that brightly  glowing  ring 
formations appear near the maximum of an ultra  soft  X-ray  radiation 
(with the energy $\leq10~keV$).  During  about  5~$ns$  these  formations 
"collapse" on the axis (see Fig. 5). Figure 6 taken from 
[19] shows 
integral pinhole photographs of X-ray  radiation.  One  can  see  that 
these are mappings of the discharge corona and cross stripes  (strata) 
which superimpose each other. Shadow regions of Fig. 6 correspond to a 
"cold" core,  along  which  a  negligible  small part 
of the current  flows.  The 
integral pinhole photographs taken by  using  filters  with  a  cutoff 
energy of 2~$keV$ show no corona glow, however one can see hot spots  of 
$\leq80~{\mu}m$ in size at the points of the axis 
corresponding to the  strata 
(Fig. 6b). The estimation of electron temperature ($T_e$) 
in the plasma corona gives 400 - 800~$eV$ and the estimates of 
the maximum temperature in 
hot spots obtained by the absorbing method are 1 - 2~$keV$ 
($T_{max} = (2.5 - 5)T_c$ ($T_c$ is the electron temperature in 
the corona), which is close 
by the order of magnitude to the estimate of our model 
(without claim of a better agreement). 
\par
It should be noted that in our model the region of a high 
temperature unlike in experiment [19] localizes 
on the axis from the very 
beginning of the process. This is due to the fact that an  undisturbed 
current is uniformly distributed throughout the cylindrical  conductor 
and not along the corona discharge as in the experiment of 
[19]. If 
the above-said is taken into account, then in terms of our model,  the 
mechanism of hot spot formation in the experiments of [19] is easily 
understood. Figure 9 of paper [2], in this 
case, corresponds to Fig. 7 
which represents a picture of hot spot formation. Subsequently, due to 
the replacement of current from the corona discharge to the core,  hot 
spots localize on the axis. 
\par
{\bf 3.6.} In the process of electric explosion  an  overheated  matter 
emerges from the near-axis zone up to  the  surface  in  the  form  of 
bright,  high-velocity  jets  which  give  rise   to   an   additional 
high-velocity shock wave in the ambient air 
[4].  These jets set a 
typical pattern (Fig. 8) where the original space period is preserved. 
Being aware of the velocity of  these  jets,  one  can  give  a  lower 
estimate for the matter temperature in the near-axis zone. For one  of 
the experiments [4] this estimation gives a 
temperature of 5~$eV$ with a 
volume average of about 0.5~$eV$.  In  the  simulation  (Fig.  4a)  the 
maximum temperature in hot spots  is  greater  by  23 times the 
average one. 
\par
{\bf 3.7.} The model enables division of the process into definite 
phases and to formalization of such widely used notions as  the  delay 
time and the switching time. In particular, in experiments 
with inductive accumulators of power  there  is  a  certain 
typical  time  when the 
current reaches its maximum. From this time the current starts 
decreasing whereas the Ohmic component of voltage  continues  increasing.  A 
section of a negative differential resistance appears on the
$UI$-characteristics. This moment is highlighted in our model 
(circuit S3 from 
[2]): a topological rearrangement of singular 
points of the dynamic 
system defined by Eqs. (6) and (7) of [2] takes place 
here.  As a result of 
this rearrangement the first system of current vortexes is formed  and 
the conductor ceases to be simply-connected (the separatrix surface of 
the  current  velocity  field  of  conduction  electrons  divides  the 
conductor into two channels (Fig. 12 of [2]). 
\par
The analysis of the X-ray  picture  series  exhibits  that  at  a 
certain moment lying to the right of  the  current  maximum  when  its 
value is on the level of 0.8 - 0.9 from the maximum, a fast process of 
the conductor destruction begins (this process is a phase of 
commutation). On 
current and voltage waveforms this process is not marked in  any  way. 
However here appears a fracture on the time dependence of the 
Kolmogorov-Sinai entropy (see Fig. 5 in [2]) 
which justifies the division of 
the process into stages: the initial and final stages of  commutation. 
There is one more typical point, namely the  one  where  the  rate  of 
entropy growth reaches its maximum. It is timed to the  amplitudes  of 
current on the level of 0.1 - 0.2 from the maximum. It  is  convenient 
to connect this moment with the end of the commutation phase. 
\par
{\bf 3.8.} While other conditions being equal, the  critical magnetic 
Rayleigh number is smaller the lower the conductivity. Therefore,  the 
non-linear effects of transition metals under discussion  have  to  be 
more strongly pronounced than those of such materials as 
silver and copper, corresponding to experimental observations. 
In  particular,  the  anomalous 
dependence of resistance on the absorbed energy in  transition  metals 
is more expressed [4]. 
\section{Conclusion} 
The results obtained in [1], [2] and discussed above are mostly of a 
qualitative character. As pointed in [1], they 
loose applicability 
from the beginning of the  space  scale  splitting.  The  sequence  of 
doubling the space scale pointed in [1]: 
$k_0{\rightarrow}{k_1} = 0.5{k_0}{\rightarrow}{k_2} = 
2{k_0}{\rightarrow}{k_3} = 2{k_2}{\rightarrow}{k_4} = 
2{k_3}{\rightarrow}\cdots$ is peculiarly analogous to a 
hypothesis 
of the scale invariance in the  second-kind  phase  transition  theory 
[22]. Since our model is the model of a 
non-equilibrium phase transition, an analog for the critical 
point is the magnetic Rayleigh 
number, where stationary states are absent if the latter  is  exceeded 
(see Fig. 3a from [1] presenting the bifurcation curve). 
\par
It appears that in order to describe the first stage of splitting 
and the formation of a low-temperature plasma with a condensed 
disperse phase (${k_0}\rightarrow{k_1}$) one can also limit 
oneself to three modes of 
perturbation [15]. It has to be noted that the 
dynamics of the process 
in this stage is qualitatively similar to the destruction of 
superconductivity by a critical current [23], 
[24] and therefore it can be 
interesting to construct a model for the destruction of a 
superconducting state as the model of a non-equilibrium phase 
transition which 
occurs as a result of the "crisis" in the two-liquid 
magneto-hydrodynamics of the superconductor. 
\par
Nevertheless, despite the limitation mentioned, we were  able  in 
terms of our model and the results of simulation to give a qualitative 
explanation from a common standpoint to experimental  results  on  the 
electric explosion of conductors,  earlier  considered  anomalous.  In 
particular, the mechanism of localizing  heat  sources,  according  to 
which the formation of hot plasma spots in  plasma-like  media  occurs 
not on account of a strong dependence of local kinetic coefficients on 
the density and temperature (the  medium  in  our  model  was  assumed 
incompressible with constant kinetic coefficients) and also 
sausage-type 
instabilities  (pinches)  but  owing  to  the  formation   of   vortex 
structures in the system (see Fig. 7 and Fig. 9 from [2]). 
\par
Since we have shown the possibility of hot spots formation  in  a 
liquid metal, it has to be expected that hot spots can exist  near  the 
cathode surface from the side of the metal and be responsible for  the 
initiation and dynamics of vacuum arc cathode spots 
[25] the behavior 
of which has been obscure so far. In our opinion,  hot  spots  play  a 
leading role in the phenomenon called the explosive electron  emission 
which is widely used in applied physics [26], [27]. 
\par
It should be emphasized  that  the  processes  discussed  in  our 
series refer to a  class  of  non-equilibrium  phase  transitions  and 
therefore the  simplest  model  proposed  by  us  (the  incompressible 
plasma-like medium with constant local kinetic  coefficients  and  the 
three space modes of perturbation) enabled a qualitative understanding 
of a wide class of experiments including the  mechanism  of  hot  spot 
formation in experiments on micro-pinches [17]. 
\ack
\par
We are pleased to thank Mr. K Ye Bobrov for assictance making the 
computer-drawings for this series.
\newpage
\references

\numrefjl{[1]}{Volkov~N~B and Iskoldsky~A~M 1993}
{J. Phys. A: Math. and Gen.}{{\rm (In this volume).}}

\numrefjl{[2]}{Volkov~N~B and Iskoldsky~A~M 1993}
{J. Phys. A: Math. and Gen.}{{\rm (In this volume).}}

\numrefjl{[3]}{Lebedev~S~V and Savvatimsky~A~I 1985}
{Sov. Phys. Usp.}{144}{243.}

\begin{sloppypar}
\numrefbk{[4]}{Iskoldsky~A~M 1985}{Thesis}{(High Current 
Electronics Institute: Tomsk) (in Russian).}
\end{sloppypar}

\begin{sloppypar}
\numrefbk{[5]}{Boortzev~V~A, Kalinin~N~V and Lootchinsky~A~V
1990}{Electrical Explosion of Conductors and its Use 
in Electrophysical Devices 
{\rm (Energoatomisdat: Moskow) (in Russian).}}
\end{sloppypar}

\begin{sloppypar}
\numrefbk{[6]}{Vichrev~V~V and Braginsky~S~I 1980 in}
{Voprosi teorii plasmi, {\rm eds. M A Leontovitch, 
vol. 10 (in Russian).}}
\end{sloppypar}

\numrefjl{[7]}{Fillipov~N~V, Fillipova~T~I and Vinogradov~V~L 
1962} {Nucl. Fusion Suppl.}{{\rm part} 2.} 

\numrefjl{[8]}{Weber~B~V at el 1987}
{IEEE Trans. Plasma Science}{PS-15}{635.}

\numrefjl{[9]}{Baikov~A~P, Iskoldsky~A~M and 
Nesterichin~Yu~E 1975}{JTF}{45}{136 (in Russian).}

\numrefjl{[10]}{Lebedev~S~V and Savvatimsky~A~I 1981}
{Tepl. vis. temp.}{19}{1184 (in Russian).}

\numrefjl{[11]}{Abramova~K~B, Zlatin~N~A and 
Peregood~B~P 1975}{JETF}{69}{2007 (in Russian).}

\numrefjl{[12]}{Bodrov~S~G, Lev~M~A and Peregood~B~P 1978}
{JTF}{48}{2519 (in Russian).}

\numrefbk{[13]}{Kadi\'c~A and Edelen~D~G~B 1983}
{A gauge theory of
dislocations and disclinations {\rm (Springer-Verlag: Berlin).}}

\numrefbk{[14]}{Lamb~H. 1932}{Hydrodynamics 
{\rm (University Press: Cambridge).}}

\numrefbk{[15]}{Volkov~N~B and Iskoldsky~A~M 1991 in}
{Proc.  8th  All-Union 
Conf. on Physics of Low-Temperature Plasma}{(Institute  of 
Physics Belorussia Academy of Science: Minsk) vol. 1 (in 
Russian).}

\numrefbk{[16]}{Frenkel~Ja~I 1975}
{The Kinetic Theory of Liquid {\rm (Nauka: Leningrad) (in 
Russian).}} 
                              
\numrefjl{[17]}{Korop~E~D, Meyerovitch~B~E, Sidelnikov~Ju~V 
and Soochorukov~O~B 1979}{Usp. fiz. nauk}{129}{87 (in Russian).}

\numrefjl{[18]}{Baksht~R~B, Datsko~I~M and Korostelev~A~F 
1985}{JTF}{55}{1540 (in Russian).}

\numrefjl{[19]}{Aivazov~I~K, Arantchook~L~E, Bogoljubsky~S~L 
and Volkov~G~S 1985}{Pis'ma v JETF}{41}{111 (in Russian).}

\numrefbk{[20]}{Baksht~R~B 1991}
{Plasmennii stolb, obrasovannii vzrivom 
microprovodnikov {\rm (Institute of High Current Electronics  
Russian  Academy of Science, Siberia Division: 
Tomsk) preprint N6 (in Russian).}}

\numrefjl{[21]}{Vichrev~V~V, Ivanov~V~V and Koshelev~K~N  1982}
{Fisika  plasmi}{8}{1211 (in Russian).}

\numrefbk{[22]}{Ma~Shang-keng 1976}{Modern Theory of Critical Phenomena 
{\rm (W. A. Benjamin, Inc.: Massachusetts).}}

\numrefbk{[23]}{London F. 1937}{Une conception 
nouvelle de la supraconductivite {\rm (Paris).}}
                  
\numrefjl{[24]}{Andreev~A~F 1968}{JETF}{54}{1510 (in Russian).}

\numrefbk{[25]}{Mesyats~G~A and Proskurovsky~D~I 1989}
{Pulsed Eletrical Discharge in Vacuum {\rm (Springer-Verlag: Berlin).}}

\numrefjl{[26]}{Boogaev~S~P, Iskoldsky~A~M, Mesyats~G~A 
and Proskoorovsky~D~I 1967}{JTF}{37}{2206 (in Russian).}
           
\numrefjl{[27]}{Foorsey~G~N, Vorontsov-Vel'yaminov~P~N 1967}
{JTF}{37}{1870 (in Russian).}
\newpage
\figures
\figcaption{Current (Fig. 1a) and voltage (Fig.  1b) waveforms
and the radius as a function of time (Fig. 1c) obtained 
during the electric explosion of a copper wire 0.4 mm 
in diameter and 10 $cm$ long.
Circuit parameters: $C = 4.2~{\mu}F$, $L = 16~nH$, 
$U = 30~kV$ [4]. 
The curve $u(t)$ (Fig. 1b) exhibits the following typical points: 
a - the start of melting ($t_a$); 
b - the end of melting ($t_b$); 
c - the onset of the explosion ($t_c$); 
d - the maximal rate of the voltage growth ($t_d$); 
e - the maximal voltage ($t_e$).}
\figcaption{Results of simulation illustrating the influence of 
the rate of energy 
input into the conductor on the dynamics of vortex 
structures and states of current. Circuit S2 from [2]. 
a - $Q(t) = \int_{0}^{l}UIdt$ is the absorbed energy; 
b - $R_{eff}(t) = UI^{-1}$ is the effective 
conductor resistance; c - $R_{eff}(Q) = UI^{-1}$. 
Curves 1 correspond to the value of a 
dimensionless resistance 
of the load $\Pi_3 = 1$; curves 2 - to $\Pi_3 =4$.}
\figcaption{X-ray photograph of the conductor cross-section 
at various times: 
"a" corresponds to the time $t_c$, "b" - $t_d$, "c" - $t_e$, 
obtained during 
the explosion of a copper wire 0.58 $mm$ in 
diameter. The circuit 
parameters: the period 40 ${\mu}s$, the capacity 
$C = 4.2~{\mu}F$ and the initial voltage $U = 30~kV$ [4].}
\figcaption{Spatial distribution of the temperature (a), the
radiation (b) and the current curve (c), obtained 
by simulation for the circuit S2 from [2]. Marked 
in the current curve is the time for the plots in Figs. 4a, b.}
\figcaption{X-ray photographs of the plasma channel obtained
in [19] by the X-ray electron-optical image converter.  Bright 
glowing points are 
ring-wise formations. The interval between the 
frames is 10 $ns$.}
\figcaption{Integral X-ray pinhole photographs obtained in
[19] in various spectral ranges.}
\figcaption{Qualitative picture of "hot" plasma  ring  
formations  in  the experiment of [19] according to our model: 
a - the distribution of the undisturbed magnetic field; 
b -  the  cross-section  of  the  separatrix  surface  of  the 
vector velocity 
field of medium particles by the plane $r - z$; 
c  -  the  cross-section  of  the  vector  velocity  field  of 
conduction 
electrons by the plane $r - z$. Marked by the 
bold point symbol is the localization of 
hot spots in Figs. 7a, b.} 
\figcaption{Optical image of the exploding conductor (a) and 
shadow X-ray pictures (b and c) obtained in [4] by using 
the electron-optical image converter and pulse shadow 
X-ray photography (the exposure time of the optical 
pattern is 5 $ns$ and of the  X-ray pattern 
- 25 $ns$). The bright glow in Fig. 8a corresponds to 
high-velocity jets formed as a result of the escape of 
overheated matter from the near-axis zone up to conductor
surface.} \newpage
\hskip 1in
\special{em:graph a3r01.pcx}
\newpage
\hskip 0.5in
\special{em:graph a3r02a.pcx}
\newpage
\hskip 0.5in
\special{em:graph a3r02b.pcx}
\newpage
\hskip 0.5in
\special{em:graph a3r02c.pcx}
\newpage
\hskip 1in
\special{em:graph a3r03.pcx}
\newpage
\hskip 1in
\special{em:graph a3r04.pcx}
\newpage
\hskip 1in
\special{em:graph a3r05.pcx}
\newpage
\hskip 1in
\special{em:graph a3r06.pcx}
\newpage
\hskip 1in
\special{em:graph a3r07.pcx}
\newpage
\hskip 1in
\special{em:graph a3r08.pcx}
\end{document}